 \documentclass[final,3p,times,twocolumn]{elsarticle}

 \usepackage{graphics}
 \usepackage{graphicx}

\usepackage{amssymb}
\usepackage{amsmath}

\journal{Journal of Molecular Spectroscopy}

\begin{document}

\begin{frontmatter}



\title{Rotational spectroscopy of oxirane-\textit{2,2}-$d_2$, $c$-CD$_2$CH$_2$O, 
       and its tentative detection toward IRAS 16293$-$2422~B}


\author[Koeln]{Holger S.P.~M\"uller\corref{cor}}
\ead{hspm@ph1.uni-koeln.de}
\cortext[cor]{Corresponding author.}
\author[Copenhagen]{Jes K. J{\o}rgensen}
\author[Rennes]{Jean-Claude Guillemin} 
\author[Koeln]{Frank Lewen} 
\author[Koeln]{Stephan Schlemmer} 

\address[Koeln]{Astrophysik/I.~Physikalisches Institut, Universit{\"a}t zu K{\"o}ln, 
  Z{\"u}lpicher Str. 77, 50937 K{\"o}ln, Germany}
\address[Copenhagen]{Niels Bohr Institute, University of Copenhagen, {\O}ster Voldgade 5$-$7, 
  1350 Copenhagen K, Denmark}
\address[Rennes]{Univ Rennes, Ecole Nationale Sup{\'e}rieure de Chimie de Rennes, 
  CNRS, ISCR$-$UMR 6226, 35000 Rennes, France}

\begin{abstract}

We prepared a sample of oxirane doubly deuterated at one C atom and studied its rotational spectrum 
in the laboratory for the first time between 120~GHz and 1094~GHz. Accurate spectroscopic parameters 
up to eighth order were determined, and the calculated rest frequencies were used to identify 
$c$-CD$_2$CH$_2$O tentatively in the interstellar medium in the Atacama Large Millimeter/submillimeter 
Array Protostellar Interferometric Line Survey (PILS) of the Class~0 protostellar system IRAS 16293$-$2422. 
The $c$-CD$_2$CH$_2$O to $c$-C$_2$H$_4$O ratio was estimated to be $\sim$0.054 with $T_{\rm rot} = 125$~K. 
This value translates to a D-to-H ratio of $\sim$0.16 per H atom which is higher by a factor of 4.5 than 
the $\sim$0.036 per H atom obtained for $c$-C$_2$H$_3$DO. Such increase in the degree of deuteration 
referenced to one H atom in multiply deuterated isotopologs compared to their singly deuterated variants 
have been observed commonly in recent years.

\end{abstract}


\end{frontmatter}




\section{Introduction}
\label{introduction}

Oxirane, $c$-C$_2$H$_4$O, is a small cyclic molecule that is also know as ethylene oxide, 
oxacyclopropane, epoxyethane, and dimethylene oxide. Here and in the following, unlabeled 
C and O atoms refer to $^{12}$C and $^{16}$O, respectively. The molecule was detected in a 
multitude of astronomical sources such as high- and low-mass star-forming regions and prestellar cores 
\cite{det-c-C2H4O_1997,more_obs_c-C2H4O_1998,still_more_obs_c-C2H4O_2001,with-O_Miguel_2008,PILS_COMs_2017,c-C2H4O_L1689B_2019}.
Its rotational spectrum was recorded in the early days of microwave spectroscopy and 
include the determination of structural parameters and of its dipole moment through 
Stark effect measurements \cite{c-C2H4O_S_rot_dip_1951}. Several further investigations 
were carried out for the main isotopolog since then 
\cite{c-C2H4O_rot_1974,c-C2H4O_FASSST_1998,c-C2H4O_FIR_2012,c-C2H4O_rot_2022}. 
The spectra of numerous isotopic species, including multiply substituted ones, were also 
recorded \cite{c-C2H4O_div-isos_rot_1974,c-C2H4O_rot_isos_1974}, albeit in the microwave 
and lower millimeter-wave regions only until quite recently. 
A sample in natural isotopic composition was used to extend the $c$-$^{13}$CCH$_4$O and 
$c$-C$_2$H$_4$$^{18}$O data sets into the upper millimeter and submillimeter regions 
\cite{c-C2H4O_rot_2022}. In addition, isotopically enriched samples were employed to 
investigate the rotational spectrum of mono-deuterated oxirane, $c$-C$_2$H$_3$DO, 
in the millimeter and far-infrared regions \cite{c-C2H3DO_rot_2019} and in the 
submillimeter region \cite{c-C2H3DO_rot_2023}. The combined data of these two studies 
and an earlier one \cite{c-C2H4O_div-isos_rot_1974} led to the detection of $c$-C$_2$H$_3$DO 
toward IRAS 16293$-$2422 \cite{c-C2H3DO_rot_2023}.

IRAS 16293$-$2422 is a Class~0 solar-type protostellar system and an astrochemical template source. 
The results on $c$-C$_2$H$_3$DO were obtained in the framework of the Protostellar Interferometric 
Line Survey (PILS), which is an unbiased molecular line survey of this source around 345~GHz 
carried out with the Atacama Large Millimeter/submillimeter Array (ALMA) to study its physical 
conditions and molecular complexity \cite{PILS_overview_2016}. Among the most prominent results 
of PILS were the detections of the organohalogen compound methyl chloride (CH$_3$Cl) 
\cite{PILS_MeCl_2017}, nitrous acid (HONO) \cite{PILS_HONO_2019}, and the tentative detection of 
3-hydroxypropenal \cite{3-hydroxypropenal_det_2022}. Several other molecules were detected for 
the first time in a low-mass star-forming region in the PILS data, among them oxirane, propanal, 
and propanone \cite{PILS_COMs_2017}, the latter is also known as acetone. 
Noteworthy are also the first interstellar detections of a plethora of isotopic species 
containing, in particular, $^{13}$C or D. Several examples and an overview of the detections 
at that time were given in \cite{PILS_div-isos_2018}; more recent examples were summarized in 
\cite{c-C2H3DO_rot_2023}.

The enrichment of deuterium in dense molecular clouds has attracted considerable interest 
for many years, as it has been viewed as an evolutionary tracer in low-mass star-forming regions 
and/or a means to trace the formation histories of complex organic molecules 
\cite{deuteration_1989,deuteration_2005,deuteration_2007,taquet14,deuteration_and_c-C3H2_2018}. 
An important aspect in this context is the observation that the degree of deuteration 
referenced to one H atom is frequently higher in multiply deuterated isotopologs than in 
the mono-deuterated isotopologs. 
Examples from PILS are D$_2$CO \cite{PILS_H2CO_2018}, CHD$_2$CN \cite{PILS_nitriles_2018}, 
CHD$_2$OCHO \citep{PILS_dideu-MeFo_2019}, CHD$_2$OCH$_3$ \cite{CH3OCHD2_rot_det_2021}, CHD$_2$OH 
\cite{CHD2OH_catalog_2022}, and CD$_3$OH \cite{CD3OH_rot_2022}. Other examples include 
D$_2$S \cite{D2S_det_2003}, HD$_2^+$ \cite{HD2+_THz_2017}, ND$_2$ \cite{NHD_ND2_det_2020} and 
D$_2$O \cite{D2O_HDO_2021}. The higher degree of deuteration per H atom for multiply deuterated 
isotopologs compared with the respective singly deuterated ones was proposed to be inherited 
from the prestellar phase. An investigation into the deuteration of thioformaldehyde of the 
prestellar core L1544 \cite{H2CS-deuteration_L1544_2022} supports this proposal. 
Additional support comes from laboratory studies of the D-to-H and H-to-D exchange of 
formaldehyde isotopologs at very low temperatures where the first reaction proceeds 
faster than the second and leads to more D$_2$CO than expected from a statistical 
distribution of D between H$_2$CO, HDCO, and D$_2$CO \cite{H-to-D-exchange_HDCO_2009}. 
The degree of deuteration in $c$-C$_2$H$_3$DO and its line intensities were sufficiently high 
in the PILS data that it appeared plausible to detect doubly deuterated oxirane as well 
\cite{c-C2H3DO_rot_2023}. While there is only one singly deuterated oxirane isotopomer, 
there are three in the case of doubly deuterated oxirane: oxirane-\textit{2,2}-$d_2$, 
oxirane-$anti$-\textit{2,3}-$d_2$, and oxirane-$syn$-\textit{2,3}-$d_2$. 
An early, limited microwave study of the last two isotopomers \cite{c-C2H4O_div-isos_rot_1974} 
has been extended recently \cite{2-3-D2_rot_2021}, but has not appeared in the literature yet 
to the best of our knowledge. No rotational data have been published for the isotopomer 
oxirane-\textit{2,2}-$d_2$. Therefore, we decided to synthesize it, analyze its rotational spectrum, 
and search for it in the PILS data.


\section{Laboratory spectroscopic details}
\label{lab-spec_details}

\begin{figure}
	\includegraphics[width=.95\columnwidth]{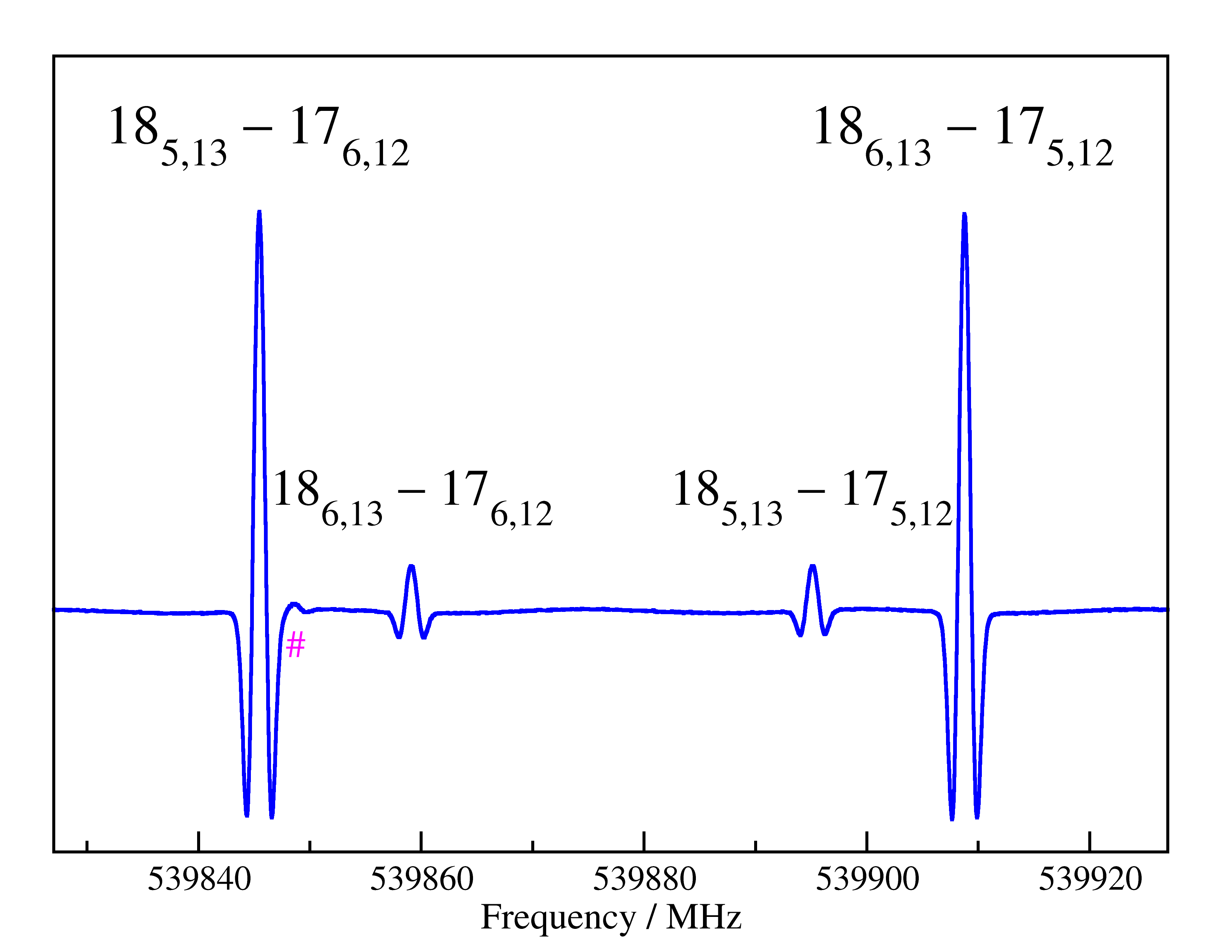}
    \caption{Section of the rotational spectrum of $c$-CD$_2$CH$_2$O. Two stronger $b$-type 
             transitions approach oblate pairing and are accompanied by weaker $a$-type 
             transitions between them. The weak line indicated by the magenta pound sign 
             is the 12$_{10,5} - 14_{10,4}$ transition of $c$-C$_2$H$_3$DO which is present 
             as an impurity.}
    \label{fig:oblate-pairing}
\end{figure}

Our measurements were carried out at room temperature using two different spectrometers. 
Pyrex glass cells with an inner diameter of 100~mm were used in both cases. Both employ 
VDI frequency multipliers driven by Rohde \& Schwarz SMF~100A microwave synthesizers 
as sources. A Schottky diode detector was employed between 120 and 181~GHz, a closed cycle 
liquid He-cooled InSb bolometer (QMC Instruments Ltd) was utilized between 490 and 
1094~GHz. We applied frequency modulation throughout to reduce baseline affects. 
The demodulation at $2f$ causes an isolated line to appear close to a second derivative 
of a Gaussian, as is displayed in Fig.~\ref{fig:oblate-pairing}.


Two connected 7~m long double pass cells equipped with Teflon lenses were utilized for 
measurements between 120 and 181~GHz. The pressures were between 1.0 and 1.5~Pa. 
Further information on this spectrometer is available elsewhere \cite{n-BuCN_rot_2012}. 
Measurements between 490 and 1094~GHz were performed using a 5~m long single pass cell at 
pressures of $\sim$0.5~Pa to around 3~Pa. Further information on this spectrometer system 
is available in \cite{CH3SH_rot_2012}.

The 490$-$750~GHz region was recorded in its entirety one time at a pressure of 1.0~Pa 
and lower detector sensitivity to avoid saturation effects on the strongest lines. 
The spectral region was recorded a second time at a pressure of 3.0~Pa with higher 
detector sensitivity and slightly higher modulation depth to optimize the signal-to-noise 
ratio (S/N) for weak lines. 
The assigned line uncertainties were dominated by the line shape; the S/N was important 
in particular for weaker symmetric lines. We assigned 5~kHz to strong, isolated, and 
very symmetric lines up to 50~kHz for weak and somewhat asymmetric lines or lines fairly 
close to other lines. This is in keeping with our investigations on $c$-C$_2$H$_4$O 
\cite{c-C2H4O_rot_2022} and $c$-C$_2$H$_3$DO \cite{c-C2H3DO_rot_2023}. 
Frequency accuracies of 10~kHz were achieved at even higher frequencies for H$_2$C$^{18}$O 
and H$_2$C$^{17}$O measured in natural isotopic composition \cite{H2CO_rot_2017} as well as 
for excited vibrational states of CH$_3$CN \cite{MeCN_up2v4eq1_etc_2021}.

Individual transitions were recorded in the other frequency windows. Strong and very strong 
lines were recorded between 973 and 1094~GHz at pressures of 1.0 and $\sim$0.5~Pa, respectively. 
The very high S/N and the very symmetric line shapes yielded uncertainties of 3 and 2~kHz, 
respectively. Weak lines with high $K_a$ and very weak lines with high $K_c$ were recorded 
between 765 and 963~GHz at a pressure of 3~Pa. The accuracies are between 10 and 50~kHz.

Finally, several stronger and some weaker transitions were recorded between 120 and 181~GHz. 
We achieved frequency accuracies of 5~kHz for the best lines for this spectrometer system 
in a study of 2-cyanobutane \cite{2-CAB_rot_2017} which is a molecule with a much richer 
rotational spectrum. The S/N was very good for all lines in the present case such that the 
line uncertainties appeared to depend solely on the line shapes. They are between 2~kHz for 
very symmetric lines to 10 or 15~kHz for less symmetric lines.


\section{Synthesis of doubly deuterated oxirane, \mbox{$c$-}CD$_2$CH$_2$O}
\label{synthesis}

\subsection{2-Chloroethanol-\textit{1,1}-$d_2$}

Sodium borodeuteride (98 atom \% D, 1.0~g, 24~mmol) and dry tetrahydrofuran (THF) (10~mL) 
were introduced under nitrogen into a three-necked flask. Dry chloroacetic acid (2.26~g, 24~mmol) 
in dry THF (10~mL) was added dropwise and evolution of hydrogen gas was observed. 
The reaction mixture was refluxed for 6~h, cooled to room temperature and then hydrolyzed with 
3\,N-D$_2$SO$_4$, made alkaline with aqueous sodium carbonate, extracted with diethyl ether, 
and dried over MgSO$_4$. The solvent was carefully removed under vacuum and the purification 
was carried out on a vacuum line (0.1~mbar) with a trap cooled to $-$60$^{\rm o}$C to selectively trap 
the 2-chloroethanol-\textit{1,1}-$d_2$. Two distillations gave 1.2~g pure product (60\% yield) with 
an isotopic purity of about 90\%.

$^1$H NMR (CDCl$_3$, 400~MHz) $\delta$ 3.09 (s, brd, 1H, OH), 3.60 (s, 2H, CH$_2$Cl). 
$^{13}$C NMR (CDCl$_3$, 100~MHz) $\delta$ 46.2 ($^1J_{\rm CH} = 149.9$~Hz (t), CH$_2$Cl), 
62.0 ($^1J_{\rm CD} = 22.0$~Hz (quint.), CD$_2$).

\subsection{Oxirane-\textit{2,2}-$d_2$}

2-Chloroethanol-\textit{1,1}-$d_2$ (1.0~g, 14.4~mmol) was vaporized in a vacuum line (0.1~mbar) 
on $t$-BuOK in powder which was heated to 90$^{\rm o}$C; similar experiments are described 
in greater detail elsewhere \cite{4synth_2019}. A trap cooled to $-$100$^{\rm o}$C removed 
selectively the impurities (mainly $t$-butanol), and oxirane-\textit{2,2}-$d_2$ was selectively 
condensed in a trap cooled at 77~K. The yield was 0.38~g or 67\%.

$^1$H NMR (CDCl$_3$, 400~MHz) $\delta$ 2.60 (m, 2H, CH$_2$). $^{13}$C NMR (CDCl$_3$, 100~MHz) 
$\delta$ 40.0 ($^1J_{\rm CD} = 26.9$~Hz (quint.), CD$_2$), 40.5 ($^1J_{\rm CH} = 175.6$~Hz (t), CH$_2$).


\begin{figure}
\begin{center}
	\includegraphics[width=.65\columnwidth]{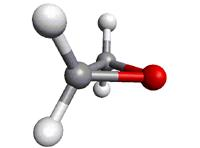}
    \caption{Model of the $c$-C$_2$H$_4$O molecule. The C atoms are shown in gray, the H atoms 
             in light gray and the O atom in red. The O atom and the mid-point of the CC bond 
             are on the $b$-axis. The two D atoms in \mbox{$c$-}CD$_2$CH$_2$O 
             are attached to the same C atom, causing a slight rotation of its inertial axis 
             system in the $ab$- (or COC) plane compared with \mbox{$c$-}C$_2$H$_4$O which 
             leads to a small $a$-dipole moment component.}
    \label{fig:molecule}
\end{center}
\end{figure}



\section{Spectroscopic properties of oxirane}
\label{properties_oxirane}

Oxirane, $c$-C$_2$H$_4$O, is a very asymmetric rotor of the oblate type with 
$\kappa = (2B - A - C)/(A - C) = 0.4093$. It has $C_{\rm 2v}$ symmetry, and the $b$-axis 
is the symmetry axis, which is determined by the O atom and the mid-point of the CC bond, 
see also Fig.~\ref{fig:molecule}. 
The four equivalent H atoms lead to \textit{para} and \textit{ortho} spin-statistics. 
The $a$-axis is parallel to the CC bond, and the $c$-axis is perpendicular to the plane 
formed by the two C atoms and the O atom. The molecule possesses a sizable $b$-dipole moment 
of 1.90~D \cite{c-C2H4O_rot_2022}; this value is 0.02~D larger than the originally reported 
one \cite{c-C2H4O_S_rot_dip_1951} because of updated values for the dipole moment of the OCS 
molecule \cite{OCS_dip_1985,OCS_dip_1986}.

Substitution of one of the four H atoms by D lowers the symmetry to $C_1$ and leads to a 
chiral species. The rotation of the inertial axes with respect to those of the main isotopolog 
introduces a small $a$-dipole moment component of 0.36~D and a very small $c$-component 
of 0.08~D, as calculated from the structure \cite{c-C2H3DO_rot_2023}. It is also an 
asymmetric rotor, but with $\kappa = +0.2042$ even further away from the oblate limit of 
$+1$. Non-trivial spin-statistics do not exist in the mono-deuterated oxirane.

Substitution of two of the four H atoms by D leads to three different isotopomers, one with 
the two D on the same C atom and two with one D on each C atom, either essentially opposite 
to each other or close to each other. The first one, oxirane-\textit{2,2}-$d_2$, is an asymmetric type 
rotor with $\kappa = +0.0229$ further away from the oblate limit still. 
The molecular symmetry is $C_{\rm s}$, and it has a calculated $a$-dipole moment component 
of 0.50~D and a $b$-component of 1.81~D. No non-trivial spin-statistics are found in this 
doubly-deuterated oxirane.
The second one is oxirane-$anti$-\textit{2,3}-$d_2$ with $C_2$ symmetry, $\kappa = +0.0838$, only a 
$b$-dipole moment component, and \textit{para} and \textit{ortho} spin-statistics exist 
in this isotopomer. The third one is oxirane-$syn$-\textit{2,3}-$d_2$ with $C_{\rm s}$ symmetry and 
$\kappa = +0.1280$. It possesses a very small $c$-dipole moment component besides an 
almost unchanged $b$-component. Non-trivial spin-statistics do not occur in this isotopomer. 
Hence, the three doubly deuterated oxirane isotopomers possess different molecular symmetries, 
which also differ from those of the mono- and undeuterated isotopologs. Please note that our 
designations of the doubly deuterated oxiranes in \cite{c-C2H3DO_rot_2023} were incorrect 
because atom 1 is the heteroatom, O in the present case, and that E and Z refer to planar 
configurations.


\section{Analysis of the $c$-CD$_2$CH$_2$O rotational spectrum and determination of 
         spectroscopic parameters}
\label{spec_analysis}


\begin{table*}
\begin{center}
\caption{Spectroscopic parameters$^a$ (MHz) of oxirane-\textit{2,2}-$d_2$ using Watson's S reduction of the rotational 
         Hamiltonian in the oblate representation III$^l$ and in the prolate representation I$^r$ along with 
         the dimensionless weighted standard deviation wrms.}
\label{tab_para_S-red}
\renewcommand{\arraystretch}{1.08}
\begin{tabular}[t]{lr@{}lr@{}lr@{}lcr@{}l}
\hline \hline
 & \multicolumn{6}{c}{III$^l$} & & \multicolumn{2}{c}{I$^r$} \\
\cline{2-7} \cline{9-10}
Parameter & \multicolumn{2}{c}{Experimental} & \multicolumn{2}{c}{B3LYP/aVTZ$^b$} &  
\multicolumn{2}{c}{Scaled$^c$} & & \multicolumn{2}{c}{Experimental} \\
\hline
$A$                      &  23062&.786666~(35) &  23018&.956    &  23066&.253      & &   23062&.777841~(35) \\
$B$                      &  18013&.451323~(29) &  17925&.922    &  18013&.056      & &   18013&.452264~(29) \\
$C$                      &  12727&.735907~(32) &  12663&.953    &  12727&.519      & &   12727&.743322~(32) \\
$D_K \times 10^3$        &     10&.18394~(7)   &     10&.117    &     10&.117      & &      12&.53427~(13)  \\
$D_{JK} \times 10^3$     &  $-$36&.85961~(12)  &  $-$36&.652    &  $-$39&.767      & &      25&.30088~(14)  \\
$D_J \times 10^3$        &     33&.52691~(8)   &     33&.286    &     33&.852      & &      12&.33667~(6)   \\
$d_1 \times 10^3$        &      7&.395157~(30) &      7&.3566   &      7&.9437     & &    $-$3&.435028~(17) \\
$d_2 \times 10^3$        &   $-$0&.927347~(11) &   $-$0&.940    &   $-$1&.0172     & &    $-$0&.692299~(7)  \\
$H_K \times 10^9$        &  $-$36&.618~(67)    &  $-$20&.28     &  $-$20&.28       & &       6&.023~(181)   \\
$H_{KJ} \times 10^9$     &     64&.224~(160)   &     37&.57     &     37&.57       & &     125&.017~(310)   \\
$H_{JK} \times 10^9$     &  $-$40&.398~(164)   &  $-$53&.7      &  $-$80&.5        & &   $-$92&.878~(194)   \\
$H_J \times 10^{9}$      &     15&.556~(75)    &     36&.6      &     36&.6        & &      14&.799~(48)    \\
$h_1 \times 10^{9}$      &   $-$7&.585~(35)    &  $-$14&.2      &   $-$1&.42       & &       4&.109~(14)    \\
$h_2 \times 10^{9}$      &     11&.005~(19)    &      5&.56     &      6&.67       & &    $-$1&.169~(8)     \\
$h_3 \times 10^{9}$      &   $-$0&.119~(4)     &   $-$2&.07     &      2&.11       & &       0&.739~(2)     \\
$L_K \times 10^{12}$     &   $-$0&.359~(25)    &       &        &       &          & &    $-$3&.628~(133)   \\
$L_{KKJ} \times 10^{12}$ &      1&.041~(72)    &       &        &       &          & &       1&.624~(246)   \\
$L_{JK} \times 10^{12}$  &   $-$1&.381~(92)    &       &        &       &          & &       2&.033~(188)   \\
$L_{JJK} \times 10^{12}$ &      1&.261~(61)    &       &        &       &          & &    $-$1&.841~(81)    \\
$L_J \times 10^{12}$     &   $-$0&.561~(21)    &       &        &       &          & &       0&.070~(14)    \\
$l_1 \times 10^{15}$     &    284&.4~(116)     &       &        &       &          & &   $-$26&.9~(37)      \\
$l_2 \times 10^{15}$     & $-$160&.7~(75)      &       &        &       &          & &   $-$58&.0~(28)      \\
$l_3 \times 10^{15}$     &    127&.1~(24)      &       &        &       &          & &       2&.37~(79)     \\
$l_4 \times 10^{15}$     &  $-$18&.8~(7)       &       &        &       &          & &    $-$1&.05~(13)     \\
wrms                     &      0&.794         &       &        &       &          & &       0&.794         \\
\hline
\end{tabular}\\[2pt]
\end{center} 
$^a$Numbers in parentheses are one standard deviation in units of the least significant figures.\\
$^b$Ground state rotational and equilibrium quartic and sextic centrifugal distortion parameters 
    calculated employing B3LYP/aug-cc-pVTZ.\\ 
$^c$B3LYP/aug-cc-pVTZ spectroscopic parameters scaled with ratios determined from experimental and 
    quantum-chemically calculated values from $c$-C$_2$H$_3$DO and $c$-C$_2$H$_4$O.\\ 
\end{table*}


\begin{table}
\begin{center}
\caption{Spectroscopic parameters$^a$ (MHz) of oxirane-\textit{2,2}-$d_2$ using Watson's A reduction of the rotational 
         Hamiltonian in the oblate representation III$^l$ and in the prolate representation I$^r$ along with 
         the dimensionless weighted standard deviation wrms.}
\label{tab_para_A-red}
\renewcommand{\arraystretch}{1.10}
\begin{tabular}[t]{lr@{}lr@{}l}
\hline \hline
Parameter & \multicolumn{2}{c}{III$^l$} & \multicolumn{2}{c}{I$^r$} \\
\hline
$A$                       &  23062&.767432~(35) & 23062&.770912~(35) \\
$B$                       &  18013&.477996~(29) & 18013&.471157~(29) \\
$C$                       &  12727&.726637~(32) & 12727&.729962~(32) \\
$\Delta_K \times 10^3$    &     19&.45745~(12)  &    19&.45707~(11)  \\
$\Delta_{JK} \times 10^3$ &     47&.98798~(17)  &    16&.99339~(11)  \\
$\Delta_J \times 10^6$    &     35&.38168~(8)   &    13&.72127~(6)   \\
$\delta_K \times 10^3$    &     11&.47547~(14)  &     8&.06040~(8)   \\
$\delta_J \times 10^3$    &      7&.39502~(3)   &     3&.43503~(2)   \\
$\Phi_K \times 10^9$      &  $-$50&.52~(45)     &    86&.32~(14)     \\
$\Phi_{KJ} \times 10^9$   &    194&.09~(70)     &  $-$1&.26~(20)     \\
$\Phi_{JK} \times 10^9$   & $-$178&.44~(34)     & $-$44&.54~(13)     \\
$\Phi_J \times 10^{9}$    &     37&.60~(8)      &    12&.47~(5)      \\
$\phi_K \times 10^9$      &   $-$1&.55~(43)     &    81&.13~(8)      \\
$\phi_{JK} \times 10^9$   &    129&.59~(25)     & $-$18&.36~(9)      \\
$\phi_J \times 10^{9}$    &   $-$7&.70~(4)      &     4&.864~(13)    \\
$L_K \times 10^{12}$      &  $-$22&.31~(82)     &  $-$5&.20~(14)     \\
$L_{KKJ} \times 10^{12}$  &     51&.13~(148)    &     4&.43~(14)     \\
$L_{JK} \times 10^{12}$   &  $-$38&.15~(83)     &      &             \\
$L_{JJK} \times 10^{12}$  &     10&.26~(20)     &  $-$2&.48~(14)     \\
$L_J \times 10^{12}$      &   $-$0&.93~(3)      &  $-$0&.055~(14)    \\
$l_K \times 10^{12}$      &  $-$17&.76~(65)     &  $-$0&.83~(10)     \\
$l_{KJ} \times 10^{12}$   &     18&.08~(37)     &      &             \\
$l_{JK} \times 10^{12}$   &   $-$4&.17~(12)     &  $-$0&.59~(3)      \\
$l_J \times 10^{12}$      &      0&.411~(12)    &  $-$0&.030~(4)     \\
wrms                      &      0&.801         &     0&.799         \\
\hline
\end{tabular}\\[2pt]
\end{center} 
$^a$Numbers in parentheses are one standard deviation in units of the least significant figures.\\
\end{table}


We used Pickett's SPFIT and SPCAT programs \cite{spfit_1991} to fit and to calculate 
transition frequencies of oxirane-\textit{2,2}-$d_2$. In the absence of experimental data, we carried 
out quantum-chemical calculations at the Regionales Rechenzentrum der Universit{\"a}t 
zu K{\"o}ln (RRZK) using the commercially available program Gaussian~09 \cite{Gaussian09E}. 
We performed B3LYP hybrid density functional calculations \cite{Becke_1993,LYP_1988}, 
utilizing the aug-cc-pVTZ (abbreviated aVTZ) basis set \cite{cc-pVXZ_1989} 
to evaluate ground state rotational and quartic and sextic equilibrium distortion 
parameters of $c$-CD$_2$CH$_2$O, $c$-C$_2$H$_3$DO, and $c$-C$_2$H$_4$O in the S reduction 
and in the III$^l$ representation. 
The results of the last two calculations and the corresponding experimental values of 
$c$-C$_2$H$_3$DO \cite{c-C2H3DO_rot_2023} and $c$-C$_2$H$_4$O \cite{c-C2H4O_rot_2022} 
were taken to estimate correction factors to be applied to the calculated spectroscopic 
parameters of $c$-CD$_2$CH$_2$O in order to reduce short-comings of the quantum-chemical 
method and those caused by the differences between equilibrium and ground state distortion 
parameters as was done, e.g., for $c$-H$_2$C$_3$O \cite{c-H2C3O_rot_2021}.

The most easily recognizable patterns in the rotational spectrum of $c$-CD$_2$CH$_2$O 
are nearly oblate paired transitions as shown in Fig.~\ref{fig:oblate-pairing}. 
These are transitions with equal $J$ and equal $K_c$ and with $K_c$ approaching $J$, 
which display small, but resolved asymmetry splitting. 
The non-zero $a$-dipole moment component leads to two weaker transitions between the 
stronger $b$-type transitions. The respective $R$-branch transitions ($\Delta J = +1$) are 
among the strong transitions in the spectrum, in particular in the submillimeter region. 
Even stronger are oblate paired transition, where the asymmetry splitting is not resolved 
anymore, and prolate paired $R$-branch transitions, with equal $J$ and equal $K_a$ and 
with $K_a$ approaching $J$. Since $\mu _c = 0$ in $c$-CD$_2$CH$_2$O, there are no 
accompanying $c$-type transitions for nearly prolate paired transitions, in contrast to 
$c$-C$_2$H$_3$DO, which possesses a very small $c$-dipole moment component.

All transitions in the 490$-$750~GHz range assigned in the first round were calculated 
somewhat higher than observed, and the deviations increased usually with $J$. This made 
the nearly oblate paired transitions with $J - K_c = 5$, of which the second group in this 
frequency range is shown in Fig.~\ref{fig:oblate-pairing}, particularly easy to identify 
because of the relatively small asymmetry splittings and the proximity to the calculated 
positions. The calculations were higher by $\sim$35~MHz for the quartet of lines near 514.6~GHz 
and 114~MHz for the single line near 717.2~GHz. The deviations were larger for transitions with 
higher $K_c$, reaching more than 210~MHz for the unsplit $J = K_c = 28 - 27$ transitions near 
719.5~GHz. Furthermore, the scatter in the deviations within the four line pattern increased 
to several tens of megahertz in the case of asymmetry splittings of several gigahertz. 
The 177 different lines, corresponding to 244 transitions because of unresolved asymmetry splitting, 
were fit with their uncertainties on average by employing rotational, quartic, and three sextic 
distortion parameters; these were $H_K$, $H_{KJ}$, and $h_2$; $h_1$ was kept fixed to its 
estimated value, whereas $H_{JK}$, $H_J$, and $h_3$ were omitted because neither was determinable 
with significance at this point, and in each case the omission of the parameter improved 
the quality of the fit compared to keeping it fixed to the initially estimated value. 
We point out that the oblate paired transitions were represented in the line lists only by 
the two strong $b$-type transitions. Adding the weaker $a$-type transitions to the list would 
only inflate the line list, but would have no effect on the parameter values or their uncertainties.

In the subsequent assignment rounds, lines with decreasing intensities were assigned. 
These included in the second round $a$-type transitions with higher $K_a$, $b$-type 
$Q$-branch transitions ($\Delta J = 0$), higher $K_a$ transitions with $\Delta J = 
\Delta K_c = -\Delta K_a = +1$ ($^pR$-branch transitions), and $b$-type transitions 
with $\Delta K_a = 3$. Later assignments include several $a$-type transitions with 
$\Delta K_a = +2$. Eventually, the line list from this frequency region consisted of 
987 different lines, corresponding to 1362 transitions. These were fit with a full set of 
distortion parameters up to eighth order with all of them significantly determined, 
albeit $L_J$ just barely.

We point out that, even though our sample of $c$\nobreakdash-CD$_2$CH$_2$O contained 
substantial amounts of $c$\nobreakdash-C$_2$H$_3$DO and even some $c$-C$_2$H$_4$O, 
line blending with other isotopic species or with excited vibrational states appeared 
to be only slightly more important than in our investigation of $c$-C$_2$H$_3$DO 
\cite{c-C2H3DO_rot_2023} for the most part, except probably for the very weak lines 
which appeared to be more affected by line blending in the present study.

The 490$-$750~GHz region was also recorded in its entirety in our investigation of 
$c$-C$_2$H$_3$DO \cite{c-C2H3DO_rot_2023}. Therefore, we checked these recordings for the 
presence of oxirane-\textit{2,2}-$d_2$ in that sample. We identified lines easily, albeit at a level 
of only $\sim$0.1\%, which is only a modest level of enrichment compared to the $\sim$0.015\% 
expected because of the naturally occurring deuterium.

Subsequently, we added transitions recorded around 1~THz. These were 103 strong or 
very strong transitions and 52 weak transitions with high $K_a$ or very weak transitions 
with high $K_c$. Many of the intended high $K_c$ transitions were too weak, heavily 
blended, or too close to neighboring lines such that only few of them were added to 
the line list. In the case of the intended high $K_a$ transitions, a smaller fraction 
of the lines was not used. Most of these lines correspond to unresolved asymmetry doublets, 
the number of different lines from this frequency region is 85. The uncertainties 
of all spectroscopic parameters were smaller after inclusion of these lines, most notably 
$L_J$ because of the strong or very strong lines with $K_c$ close to or equal to $J$.

Finally, we included 178 stronger and 39 weaker transitions from the 120$-$181~GHz region to 
the line list. Each of them corresponds to a single transition. The parameter uncertainties 
improved somewhat as a consequence.

These data add up to 1289 different transition frequencies and 1734 transitions. 
The $J$ quantum numbers range from 2 to 58 with $K_a \le 42$ and $K_c \le 53$. 
After having determined the final line list and the final set of spectroscopic parameters 
in the S reduction and the III$^l$ representation, we also evaluated parameters applying 
the A reduction and the S and A reductions in the I$^r$ representation. The experimental 
S reduction parameters are presented in Table~\ref{tab_para_S-red} together with 
quantum-chemically calculated and scaled values in the III$^l$ representation; 
the experimental A reduction values are given in Table~\ref{tab_para_A-red}. 
The weighted standard deviations for each fit are also included in the tables. 
We also determined the weighted standard deviations of the lines from each assignment round. 
These values were between 0.64 and 0.97, mostly closer to the value of the entire fit. 
We also mention that the rms values for the four fits are all very close to 12~kHz. 
The rms value of a fit is a meaningful measure of the quality of the fit if the uncertainties 
of the lines differ very little. There is an uncertainty scatter in the present line list 
such that the rms value is dominated by the lines with larger uncertainties. 
Nevertheless, the rms values provide an indication of the average quality of the lines.

The final $c$-CD$_2$CH$_2$O line file, the parameter and fit files in the S reduction 
and the III$^l$ representation are available as supplementary material to this article. 
The line file and all parameter and fit files along with auxiliary files are provided 
in the data 
section\footnote{https://cdms.astro.uni-koeln.de/classic/predictions/daten/Oxiran/}
of the Cologne Database for Molecular Spectroscopy, CDMS \cite{CDMS_2005,CDMS_2016}. 
A calculation of the rotational spectrum of $c$\nobreakdash-CD$_2$CH$_2$O based on the fit 
in the S reduction and the III$^l$ representation is available in the CDMS 
catalog\footnote{https://cdms.astro.uni-koeln.de/classic/entries/}.


\section{Discussion of the laboratory results}
\label{lab_discussion}

The $c$-CD$_2$CH$_2$O parameter set in the S reduction and the III$^l$ representation 
is very similar in extent and absolute uncertainties to those of $c$-C$_2$H$_3$DO 
\cite{c-C2H3DO_rot_2023} and $c$\nobreakdash-C$_2$H$_4$O \cite{c-C2H4O_rot_2022}. 
Therefore, the data should be sufficiently accurate for all radio astronomical observations.

It is noteworthy that the A reduction in the III$^l$ representation yields a fit 
of essentially identical quality with the same number of spectroscopic parameters. 
This is in sharp contrast to $c$-C$_2$H$_4$O, for which a fit employing this 
combination of reduction and representation gave a still poor fit with a much 
larger parameter set. We carried out a corresponding fit for $c$-C$_2$H$_3$DO 
in order to gain more insight into the performance of the A reduction in the 
III$^l$ representation. It turned out that the same set of parameters as the 
A reduction in the I$^r$ representation yielded a fit with only slightly larger 
weighted rms error. This fit and a fit using the S reduction in the I$^r$ 
representation are available in the data section$^1$ of the CDMS \cite{CDMS_2005,CDMS_2016}. 
The poor performance of the A reduction in an oblate representation is quite common, 
see \cite{c-C2H4O_rot_2022} for several examples, but apparently not ubiquitous. 
Rotational spectra of molecules closer to the oblate limit are obviously more likely 
to yield poor fits, but $\kappa = 0.4093$ in the case of $c$-C$_2$H$_4$O is quite 
far from the oblate limit of +1, suggesting other aspects matter as well, which could, 
for example, be the quantum number range or the extent to which asymmetry plays a role 
in the experimental line list.

The rotational and quartic centrifugal distortion parameters from a quantum-chemical 
calculation agree quite well after scaling with the experimental ones, as can be seen 
in Table~\ref{tab_para_S-red}, but the agreement is less good for the sextic distortion 
parameters and in some case, the unscaled values are closer to the experimental values  
of the quartic and sextic distortion parameters. A similar procedure was applied to 
the spectroscopic parameters up to sixth order of $c$-C$_2$H$_3$DO using higher level 
coupled cluster calculations \cite{c-C2H2DO_calc_param_2014}. The sextic distortion 
parameters agree somewhat better with the experimental ones \cite{c-C2H3DO_rot_2023} 
than in the present case of $c$-CD$_2$CH$_2$O, but the quality differences are less 
pronounced for the lower order parameters. This demonstrates that lower level calculations, 
such as B3LYP, can be an alternative to coupled cluster calculations which can get 
quite demanding in terms of time and memory for larger molecules.


\section{Astronomical search toward IRAS~16293-2422~B}
\label{details-obs}

We used the calculation of the rotational spectrum of $c$-CD$_2$CH$_2$O as described 
at the end of Section~\ref{spec_analysis} to search for this isotopolog in data from 
PILS.  As mentioned in the introduction, PILS is an unbiased molecular line survey 
of the protostellar object IRAS 16293$-$2422 carried out with ALMA in its Cycle~2 
(project id: 2013.1.00278.S, PI: J.~K.~J{\o}rgensen).  The source is separated into 
two main components, the more prominent Source~A, which itself is a binary source 
\cite{wootten89}, and the secondary Source~B, separated by about 5~arcsec (700~au); 
both are commonly viewed as Class~0 protostars. The survey covers part of Band~7 
from 329.1 to 362.9~GHz at 0.2~km~s$^{-1}$ spectral and $\sim$0.5~arcsec angular 
resolution. J{\o}rgensen et al. provides an overview of the data and their reduction 
along with first results from the survey \cite{PILS_overview_2016}.


\begin{figure*}
	\includegraphics[width=1.9\columnwidth]{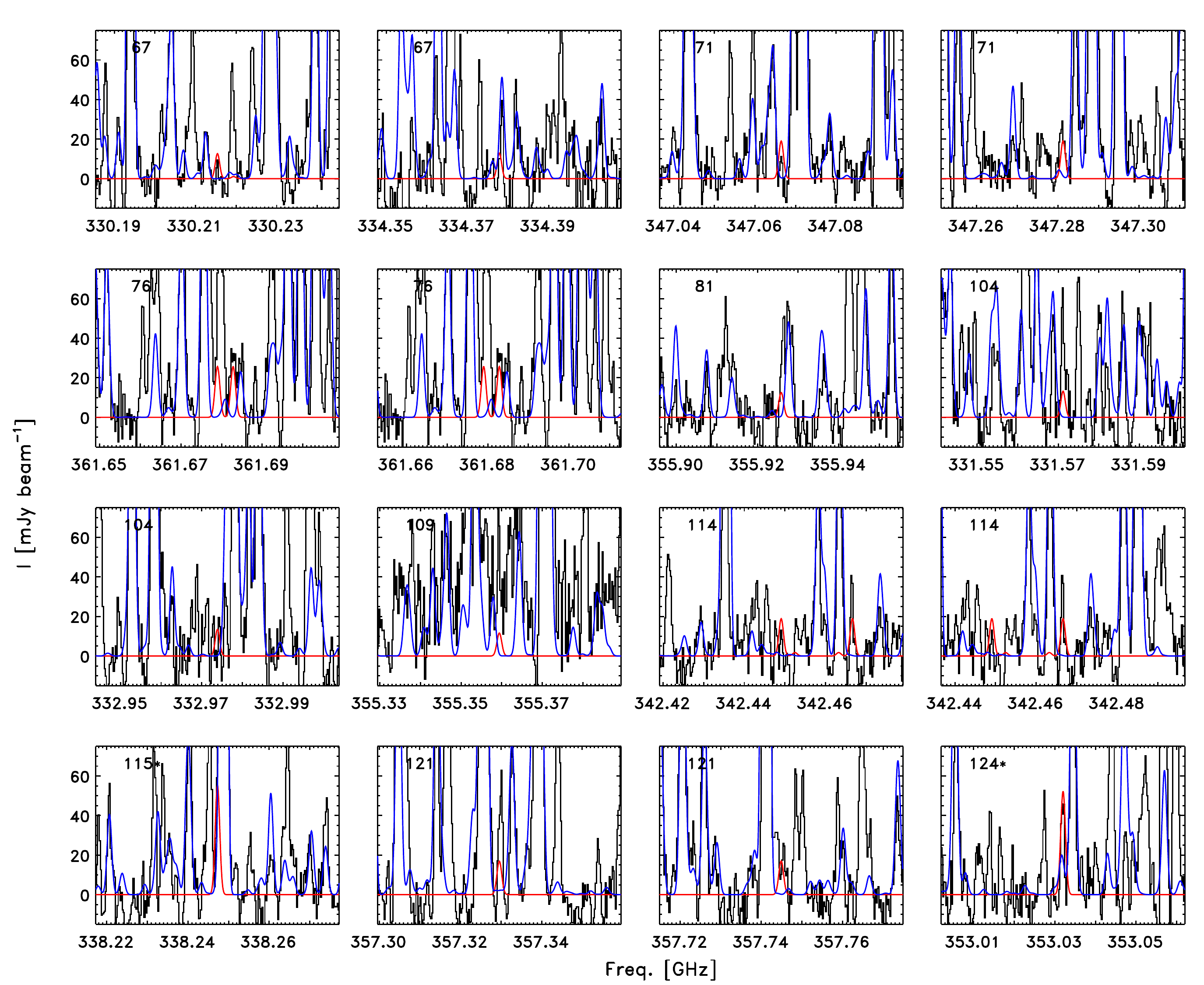}\\
    \caption{Sections of the Band~7 PILS data slightly offset from the continuum peak 
             displaying stronger lines of $c$-CH$_2$CD$_2$O. The observed data are shown 
             in black, the modeled $c$-CH$_2$CD$_2$O lines in red, and the models of all 
             other assigned lines in blue. The upper state energy of each line is given 
             in the upper left corner of each panel.}
    \label{fig:astro-spec}
\end{figure*}

\begin{table}
\begin{center}
  \caption{Sixteen emission lines (GHz) predicted to be the brightest in the PILS 
    frequency range with quantum numbers QN, upper state energy, $E_{\rm up}$ (K), 
    Einstein $A$ value ($10^{-4}$~s$^{-1}$), and line strength $W$ (mJy~km~s$^{-1}$).}
  \label{table:obslines}
\begin{tabular}{lcrcl}\hline\hline
Frequency &           QN         & $E_{\rm up}$ & $A$  & $W$ \\ 
\hline
 330.2154 & $ 8_{6,  3}- 7_{5, 2}$        &  67 &  3.52 & 15.8 \\
 331.5712 & $11_{3,  8}-10_{4, 7}$        & 104 &  3.65 & 16.5 \\
 332.9739 & $11_{4,  8}-10_{3, 7}$        & 104 &  3.71 & 16.8 \\
 334.3779 & $ 8_{6,  2}- 7_{5, 3}$        &  67 &  3.59 & 16.1 \\
 338.2473 & $13_{x, 13}-12_{x, 12}$$^{a}$ & 115 & 14.52 & 32.0 \\
 342.4491 & $12_{2, 10}-11_{3, 9}$        & 114 &  5.23 & 23.6 \\
 342.4666 & $12_{3, 10}-11_{2, 9}$        & 114 &  5.23 & 23.6 \\
 347.0665 & $ 8_{7,  2}- 7_{6, 1}$        &  71 &  5.49 & 23.8 \\
 347.2813 & $ 8_{7,  1}- 7_{6, 2}$        &  71 &  5.50 & 23.8 \\
 353.0320 & $13_{x, 12}-12_{x, 11}$$^{b}$ & 124 & 14.72 & 30.2 \\
 355.3594 & $11_{5,  7}-10_{4, 6}$        & 109 &  3.33 & 14.5 \\
 355.9258 & $ 9_{6,  4}- 8_{5, 3}$        &  81 &  3.48 & 15.7 \\
 357.3296 & $12_{3,  9}-11_{4, 8}$        & 121 &  4.94 & 21.2 \\
 357.7449 & $12_{4,  9}-11_{3, 8}$        & 121 &  4.96 & 21.3 \\
 361.6790 & $ 8_{8,  1}- 7_{7, 0}$        &  76 &  7.73 & 32.1 \\
 361.6828 & $ 8_{8,  0}- 7_{7, 1}$        &  76 &  7.73 & 32.1 \\
\hline
\end{tabular}\\[2pt]
\end{center} 
$^{a}$Blend of the oblate paired $13_{0,13}-12_{1,12}$ and $13_{1,13}-12_{0,12}$ transitions 
  (additional minor contributions from the corresponding $a$-type transitions). 
$^{b}$Blend of the oblate paired $13_{1,13}-12_{2,12}$ and $13_{2,13}-12_{1,12}$ transitions 
  (additional minor contributions from the corresponding $a$-type transitions).
\end{table}


For the search for $c$-CD$_2$CH$_2$O, we focus on a position offset by one beam 
from source~B, also targeted in several previous studies of the molecular complexity 
in PILS (e.g., \cite{PILS_overview_2016,PILS_div-isos_2018,PILS_nitriles_2018}). 
This position is ideal because of the narrow lines seen there ($\sim$1~km~s$^{-1}$), 
which allows for searches for rare species and limited continuum opacity causing 
absorption to be less severe. We performed the search by fitting synthetic spectra 
for $c$\nobreakdash-CD$_2$CH$_2$O to the ALMA data. 
The synthetic spectra are calculated under the assumption of the emission being 
optically thin and the excitation being in local thermodynamical equilibrium. 
This is a valid assumption given the high densities in the molecular cloud 
surrounding the protostar that is probed by our ALMA observations 
\citep{PILS_overview_2016}. We assumed $T_{\rm rot} = 125$~K in our initial 
calculations as was determined for $c$-C$_2$H$_4$O \cite{PILS_COMs_2017} 
and employed a linewidth (FWHM) of 1~km~s$^{-1}$ and a velocity offset relative 
to the local standard of rest of 2.6~km~s$^{-1}$, similar to other species 
toward this position.

Fig.~\ref{fig:astro-spec} shows the 16 emission lines predicted to be brightest 
for the assumed value of $T_{\rm rot}$. These represent 18 rotational transitions 
but where two sets of two transitions, respectively, are so close in frequency 
that they cannot be separated in the observations. 
The transitions, their frequencies, energies of their upper levels above 
the ground state and predicted strengths assuming the maximum possible 
column density are summarized in Table~\ref{table:obslines}. 
For some of these predicted lines, no clear signal is seen down to the RMS 
noise level of the data (1$\sigma$ being 4$-$5~mJy~beam$^{-1}$~km~s$^{-1}$), 
but a handful of predicted lines are seen to match observed features with 
line strengths above 20~mJy~beam$^{-1}$~km~s$^{-1}$. Importantly, if 
the column density is constrained based on these transitions, no lines are 
predicted to be significant where no observed emission is seen $-$ an important 
criterion for the assignment of species in astronomical spectra.

Another test of the possible assignment can be taken from the derived 
column density from the tentative lines: the line strengths of those imply 
a column density of $3.3\times 10^{14}$~cm$^{-2}$. This value can be compared 
to the column density of the mono-deuterated variant for the same $T_{\rm rot}$ 
of $8.9\times 10^{14}$~cm$^{-2}$ \cite{c-C2H3DO_rot_2023}, corresponding to 
a ratio of the doubly to mono-deuterated variants of 0.37. 
Taking $6.1\times 10^{15}$~cm$^{-2}$ for $c$-C$_2$H$_4$O \cite{PILS_COMs_2017}, 
we obtain a respective ratio of 0.054. Considering that there are statistically 
two ways to form $c$-CD$_2$CH$_2$O, but only one for $c$-C$_2$H$_4$O, this gives 
a D/H ratio of $\sim$16\% per H atom. This would, in turn, correspond to an 
enhancement by factor of 4.5 compared to the $\sim$0.036 per H atom obtained from 
the ratio between $c$-C$_2$H$_3$DO and the non-deuterated variant $c$-C$_2$H$_4$O. 
This enhancement is similar to enhancement of other di-deuterated variants 
observed toward IRAS~16293-2422~B including CHD$_2$CN \cite{PILS_nitriles_2018}, 
CHD$_2$OCHO \cite{PILS_dideu-MeFo_2019}, CH$_3$OCHD$_2$ \cite{CH3OCHD2_rot_det_2021}, 
CHD$_2$OH \cite{CHD2OH_catalog_2022}, and CHD$_2$CHO \cite{CHD2CHO_rot_det_2023}, 
and the inferred D/H ratios falls to those in the range of approximately 15$-$25\%. 
This points to a common formation mechanism of these species including efficient 
H-D substitution reactions on grain-surfaces at low temperatures with these being 
more efficient one the first deuteration has occurred, see discussion in 
\cite{CHD2OH_catalog_2022}.

Still, our astronomical results on $c$-CD$_2$CH$_2$O should be viewed with some caution 
as the number of identified lines is on the low side to claim a definitive detection. 
One way to solidify the results lies in the detection of the other two doubly deuterated 
isotopomers of oxirane. All H atoms in $c$-C$_2$H$_4$O are chemically equivalent even 
if the three doubly deuterated isotopomers display different molecular symmetries. 
We expect in particular that all doubly deuterated isotopomers display 
the same degree of deuterium enrichment. 
Thus, characterizations of those would make it possible to verify the assignment of 
$c$-CD$_2$CH$_2$O and the interpretation that it indeed shows equivalent properties 
in terms of the deuterium enhancement as the other complex organics mentioned above.


\section{Conclusion}
\label{conclusion}

We prepared a sample of oxirane-\textit{2,2}-$d_2$, $c$\nobreakdash-CD$_2$CH$_2$O, 
and analyzed its rotational spectrum for the first time. The resulting extensive set 
of spectroscopic parameters is accurate enough to search for this isotopolog in space. 
Using PILS data of IRAS 16293$-$2422, we tentatively identified a handful of lines of 
this isotopolog toward source B: with a column density derived on basis of these lines, 
no other non-detected features are predicted. The inferred column density of 
oxirane-\textit{2,2}-$d_2$ would suggest an enhancement of this di-deuterated variant 
relative to the mono-deuterated variant similar to what has been observed for other complex 
organic molecules, an indication of their common formation of low temperatures. 
We suggest further characterizations and searches of the di-deuterated isotopomers of 
oxirane to verify these results.


\section*{CRediT authorship contribution statement}

Holger S.P. M\"uller: Conceptualization, Investigation, Methodology, Formal analysis, 
Validation, Data curation, Writing $-$ Original Draft, Writing $-$ review \& editing. 
Jes K. J{\o}rgensen: Resources, Writing $-$ Original Draft, Writing $-$ review \& editing. 
Jean-Claude Guillemin: Resources, Writing $-$ Original Draft, Writing $-$ review \& editing. 
Frank Lewen: Resources, Writing $-$ review \& editing. 
Stephan Schlemmer: Funding acquisition, Resources, Writing $-$ review \& editing.

\section*{Declaration of competing interest}

The authors declare that they have no known competing financial interests or personal 
relationships that could have appeared to influence the work reported in this paper.

\section*{Data availability}

The spectroscopic line lists and associated files are available as supplementary material 
through the journal and in the data section of the CDMS.$^1$ The underlying original 
spectral recordings will be shared on reasonable request to the corresponding author. 
The radio astronomical data are available through the ALMA 
archive.\footnote{https://almascience.eso.org/aq/}


\section*{Acknowledgments}

We dedicate this work to the memories of Per Jensen and Helge Willner, 
two former professors at Bergische Universit{\"a}t Wuppertal, Germany, 
who recently passed away.  This paper makes use of the following ALMA 
data: ADS/JAO.ALMA$\#$2013.1.00278.S. ALMA is a partnership of ESO 
(representing its member states), NSF (USA) and NINS (Japan), together 
with NRC (Canada), NSC and ASIAA (Taiwan) and KASI (Republic of Korea), 
in cooperation with the Republic of Chile. The Joint ALMA Observatory 
is operated by ESO, AUI/NRAO and NAOJ.  The work in K{\"o}ln was supported by 
the Deutsche Forschungsgemeinschaft through the collaborative research center 
SFB~956 (project ID 184018867) project B3 and through the Ger{\"a}tezentrum 
SCHL~341/15-1 (``Cologne Center for Terahertz Spectroscopy''). 
We thank the Regionales Rechenzentrum der Universit\"at zu K\"oln (RRZK) for 
providing computing time on the DFG funded High Performance Computing System CHEOPS. 
J.K.J. acknowledges support from the Independent Research Fund Denmark 
(grant number 0135-00123B). J.-C. G. acknowledges support by the Centre National d'Etudes 
Spatiales (CNES; grant number 4500065585) and by the Programme National Physique 
et Chimie du Milieu Interstellaire (PCMI) of CNRS/INSU with INC/INP co-funded by 
CEA and CNES.  Our research benefited from NASA's Astrophysics Data System (ADS).

\appendix
\section*{Appendix A. Supplementary Material}

Supplementary data associated with this article can be found, in the online version, 
at https://doi.org/10.1016/j.jms.2023.111777. 
This material consists of fit, line, and parameter files along with an explanatory file. 
All files are ascii text files which are gathered in a zip file.



\bibliographystyle{elsarticle-num}
\bibliography{22D2}






\end{document}